\documentstyle[preprint,aps]{revtex}

\title{Chiral Symmetry Breaking in QED in a Magnetic Field
at Finite Temperature}

\author{V.P.~Gusynin and I.A.~Shovkovy}

\address{ Bogolyubov Institute for Theoretical Physics\\
252143, Kiev, Ukraine}

\draft
\begin{document}
\maketitle

\begin{abstract}
The catalysis of chiral symmetry breaking in the massless weakly 
coupled QED in a magnetic field at finite temperature is studied. 
The temperature of the symmetry restoration is estimated 
analytically as $T_c\approx m_{dyn} $, where $m_{dyn}$ is 
the dynamical mass of a fermion at zero temperature.  
\end{abstract}


Recently it has been shown that a constant magnetic field leads to 
a dynamical breaking of chiral symmetry at the weakest attractive 
interaction between fermions \cite{GMS1}. This effect is universal 
and is due to the dimensional reduction $D\rightarrow D-2$ in the 
fermion pairing in a strong magnetic field when the dynamics of the 
lowest Landau level (LLL) plays the crucial role.  As concrete 
models, the Nambu-Jona-Lasinio (NJL) model as well as QED in 2+1 
and 3+1 dimensions were considered 
\cite{GMS1,Klim,GMS2,GMS3,QED3,QED4,NP,Ng1,Hong}.  The effect of 
catalyzing the dynamical chiral symmetry breaking under the 
influence of a magnetic field was extended also to the case of 
external non-abelian chromomagnetic fields and finite temperatures 
\cite{Igor,Ebert,Smilga}, as well as to the supersymmetric NJL 
model \cite{Elias}, confirming the universality of the mechanism.

An important question of the chiral symmetry restoration in
QED$_4$ at finite temperature was addressed in \cite{Ng2}. Lee,
Leung, and Ng have obtained for the critical temperature $T_c\simeq
\frac{\alpha}{\pi^2}\sqrt{2\pi|eB|}$, where $\alpha$ is the fine
structure constant and $B$ is the magnetic field strength. However,
their $T_c$ can be considered only as a rough upper estimate.  In
this paper we reconsider the problem and show that the correct
estimate for the critical temperature  is $T_c\approx m_{dyn}$,
where $m_{dyn}$ is the dynamical mass of a fermion at zero
temperature which in its turn is given by $m_{dyn}\approx\sqrt
{e|B|}\exp\left(-\sqrt{\pi/\alpha}\right)$ \cite{QED4}.

The Lagrangian density of massless QED$_4$ in a magnetic field is
\begin{equation}
{\cal L}= -{1\over 4}F^{\mu\nu}F_{\mu\nu}
+ {1\over2}[\bar\psi,i\gamma^\mu D_\mu\psi],
\label{density}
\end{equation}
where the covariant derivative $D_\mu $ is
\begin{eqnarray}
D_\mu=\partial_\mu-ie(A_\mu^{ext}+A_\mu),\quad
A_\mu^{ext}=\left(0,-{B\over2}x_2,{B\over2}x_1,0\right).
\end{eqnarray}
The Lagrangian density (\ref{density}) is chiral invariant (we do 
not discuss the anomaly connected with the current $j_{5\mu}$, in 
any case it is not manifested in quenched approximation dealt with 
in this paper). As is known, there is no spontaneous chiral 
symmetry breaking at $B=0$ in the weak coupling phase of QED 
\cite{RNC}.  However, the magnetic field changes the situation 
drastically: at $B\neq0$ the chiral symmetry is broken and there 
appears a gapless Nambu-Goldstone (NG) boson composed from a 
fermion and an antifermion. The dynamical mass (energy gap) for a 
fermion can be defined by considering the Bethe-Salpeter (BS) 
equation for NG boson \cite{QED4} or the Schwinger-Dyson (SD) 
equation for the dynamical mass function \cite{Ng1,Hong}. Both 
these approaches lead to the following equation for the dynamically 
generated fermion mass in the quenched approximation (taking into 
account the dominance of LLL at strong magnetic field) and in the 
Feynman gauge:
\begin{eqnarray}
\Sigma(p_{||})=\frac{\alpha}{2\pi^2i}\int\frac{d^2k_{||}
\Sigma(k_{||})}{k^2_{||}
-\Sigma^2(k_{||})}\int_0^\infty\frac{dk^2_{\perp}
\exp(-k_\perp^2\ell^2/2)}{(k_{||}
-p_{||})^2-k_\perp^2},
\label{maineq}
\end{eqnarray}
where $p_{||}$ is a two-dimensional momentum, $p_{||}=(p_0,p_3)$
(henceforth we will omit the subscript $||$ in $p$ and $k$),
$\ell=1/ \sqrt{e|B|}$ is the magnetic length.  
In Euclidean region, the equation (\ref{maineq}) with the
replacement $\Sigma^2(k) \rightarrow \Sigma^2(0) \equiv m^2_{dyn}$
in the denominator
\begin{eqnarray}
\Sigma(p)=\frac{\alpha}{2\pi^2}\int\frac{d^2k\Sigma(k)}
{k^2+m^2_{dyn}}\int_0^\infty\frac{dx\exp(-x\ell^2/2)}{({\bf
k}-{\bf p})^2+x},
\label{mainEuc}
\end{eqnarray}
has been analyzed in \cite{NP,Hong}.  Particularly, as was shown in
\cite{NP} (see Appendix C), in the case of weak coupling $\alpha$,
the mass function $\Sigma(p)$ remains almost constant in the range
of momenta $0<p^2\alt 1/\ell^2$ and decays like $1/p^2$ outside
that region.  To get an estimate for $m_{dyn}$ at $\alpha\ll
1$, we set the external momentum to be zero and notice that the
main contribution of the integral is formed in the infrared region
with $k^2\alt 1/\ell^2$. The latter validates in its turn the
substitution $\Sigma(k) \rightarrow \Sigma(0)$ in the integrand
of (\ref{mainEuc}), and finally we come to the following gap
equation
\begin{eqnarray}
\Sigma(0)\simeq\frac{\alpha}{2\pi^2}\Sigma(0)
\int\frac{d^2k}{k^2+m^2_{dyn}}
\int_0^\infty\frac{dx\exp(-x\ell^2/2)}{k^2+x},
\label{approx}
\end{eqnarray}
i.e.
\begin{eqnarray}
1\simeq\frac{\alpha}{2\pi}\int_0^\infty
\frac{dx\exp(-ax)}{x-1}\log x,
\quad
a\equiv \frac{m^2_{dyn}\ell^2}{2}.
\label{gapeq}
\end{eqnarray}
The main contribution in (\ref{gapeq}) comes from the region
$x\alt 1/a$, thus at $a\ll 1$ we get
\begin{equation}
1\simeq \frac{\alpha}{4\pi}\log^2
\left(\frac{m^2_{dyn}\ell^2}{2}\right),
\label{HFeq}
\end{equation}
\begin{equation}
m_{dyn}\simeq C\sqrt{e|B|}\exp\left[-\sqrt{\pi\over\alpha}\right],
\label{massdyn}
\end{equation}
where $C$ is a constant of order one. The exponential factor 
displays the nonperturbative nature of this result. We note that 
the double logarithmic asymptotics of the electron self-energy in a 
magnetic field in the standard perturbation theory was obtained in 
earlier papers \cite{Jancov,Loskutov}.

More accurate analysis which takes into account
the momentum dependence of the mass function leads to the result
\cite{NP}
\begin{equation}
m_{dyn}\simeq C\sqrt{e|B|}\exp\left[-{\pi\over2}
\sqrt{\frac{\pi}{2\alpha}}\right].
\end{equation}
Notice that the ratio of the powers of this exponent and that in
Eq.(\ref {massdyn}) is $\pi/2\sqrt{2}\simeq 1.1$, thus the
approximation used above is rather reliable.

To study chiral symmetry breaking in an external field at nonzero
temperatures we use the imaginary-time formalism \cite{Abr}.  Now
the analogue of the equation (\ref{maineq}) (with the replacement
$\Sigma^2(\omega_{n},k) \rightarrow m^2(T)$ in the denominator)
reads
\begin{eqnarray}
\Sigma(\omega_{n'},{\bf p})=\frac{\alpha}{\pi}
T\sum_{n=-\infty}^\infty \int\limits_{-\infty}^{\infty}
\frac{dk\Sigma(\omega_{n},k)} {\omega^2_{n}+k^2+m^2(T)}
\int\limits^\infty_0
\frac{dx\exp(-x\ell^2/2)}{(\omega_{n}-\omega_{n'})^2+(k-p)^2+x},
\label{eq:our1}
\end{eqnarray}
where $\omega_n=\pi T(2n+1)$.

If we take now $n'=0, p=0$ in the left hand side of
Eq.(\ref{eq:our1}) and put $\Sigma (\omega_{n},k) \approx
\Sigma(\omega_{0},0) = const$ in the integrand, we come to the
equation
\begin{eqnarray}
1&=&\frac{\alpha}{\pi}T\sum_{n=-\infty}^\infty
\int\limits_{-\infty}^{\infty}
\frac{dk}{\omega_n^2+k^2+m^2(T)} \int\limits^\infty_0
\frac{dx\exp(-x\ell^2/2)}{(\omega_n-\omega_{0})^2+k^2+x}.
\label{eq:our2}
\end{eqnarray}
It is easy to check that the gap equation (\ref{eq:our2})
coincides with the equation (58) in \cite{Ng2}.

To evaluate the sum in (\ref{eq:our2}), we transform it into the
integral in complex plane $\omega$:
\begin{eqnarray}
T \sum_{n=-\infty}^\infty  \frac{1} {(\omega_n^2+a^2)
[(\omega_n-\omega_{0})^2+b^2]}
= \frac{1}{2\pi i}\int\limits_{C}
\frac{d \omega}{\left[1+e^{-\omega\beta}\right]}
\frac{1}{(\omega^2-a^2)[(\omega-i\omega_0)^2 - b^2]},
\label{eq:sum-int}
\end{eqnarray}
where $\beta\equiv 1/T$ and the contour $C$ runs as usually around 
the poles of the function $(1+e^{-\omega\beta})^{-1}$.  By 
deforming the contour, we can represent the integral in 
(\ref{eq:sum-int}) as the sum over four residues at $\omega=\pm a$ 
and $\omega=i\omega_0\pm b$.  Thus, we obtain:
\begin{eqnarray}
T \sum_{n=-\infty}^\infty  
\frac{1}{(\omega_n^2+a^2)[(\omega_n-\omega_{0})^2+b^2]} &=&
\frac{ (\pi T)^2 +b^2 -a^2 }{[(\pi T)^2 +b^2 -a^2]^2 
+ (2\pi T a)^2} \frac{\tanh(a/2T)}{2a}\nonumber\\
&+&\frac{ (\pi T)^2 +a^2 -b^2 }{[(\pi T)^2 +a^2 -b^2]^2
+ (2\pi T b)^2} \frac{\coth(b/2T)}{2b}.
\label{eq:sum}
\end{eqnarray}
Eq.(\ref{eq:sum}) can be obtained also by means of
direct summation using the formula
\begin{eqnarray}
\sum_{n=0}^\infty\frac{1}{(n+a)(n+b)(n+c)(n+d)}
&=&-\frac{\psi(a)}{(b-a)(c-a)(d-a)}
-\frac{\psi(b)}{(a-b)(c-b)(d-b)} \nonumber\\
& &-\frac{\psi(c)}{(a-c)(b-c)(d-c)}
-\frac{\psi(d)}{(a-d)(b-d)(c-d)},
\end{eqnarray}
where $\psi(x)=d\log\Gamma(x)/dx$.

Substituting Eq.(\ref{eq:sum}) with $a^2=k^2+m^2(T)$ and 
$b^2=k^2+x$ in (\ref{eq:our2}), we get the following  
equation:
\begin{eqnarray}
&&1=\frac{\alpha}{\pi} \int\limits_{0}^{\infty}
\int\limits^\infty_0
\frac{dk dx \exp[-x\ell^2/2]}{[(\pi T)^2 +x -m^2(T)]^2 
+(2\pi T)^2 (k^2+m^2(T))}
\label{eq:our3}\\
&&\times\left\{
\frac{(\pi T)^2 +x -m^2(T)}{\sqrt{k^2 +m^2(T)}}
\tanh \left(\frac{\sqrt{k^2+m^2(T)}}{2T}\right)
+\frac{(\pi T)^2 +m^2(T) -x}{\sqrt{k^2+x}}
\coth \left(\frac{\sqrt{k^2+x}}{2T}\right)
\right\}.\nonumber
\end{eqnarray}
In the limit $T\to0$, Eq.(\ref{eq:our3}) reduces to the 
following one
\begin{eqnarray}
&&1=\frac{\alpha}{\pi} \int\limits_{0}^{\infty}
\int\limits^\infty_0
\frac{dk dx \exp[-x\ell^2/2]}{\sqrt{k^2+x}\sqrt{k^2+m^2_{dyn}}
\left(\sqrt{k^2+x}+\sqrt{k^2+m^2_{dyn}}\right)},
\label{T=0}
\end{eqnarray}
which is just what one obtains from Eq.(\ref{approx}) after
performing the integration over $k_4=-ik_0$.

The equation for the critical temperature is obtained from
(\ref{eq:our3}) puting $m(T_c)=0$:
\begin{eqnarray}
1=\frac{\alpha}{\pi}
\int\limits_{0}^{\infty} \int\limits^\infty_0
\frac{dk dx e^{-2x(\pi T_c\ell)^2}}{[1/4 +x]^2 +k^2}
\left\{ \frac{1/4 +x}{k} \tanh \left(\pi k\right)
+\frac{1/4 -x}{\sqrt{k^2+x}} \coth \left(\pi\sqrt{k^2+x}\right)
\right\},
\label{eq:t_c}
\end{eqnarray}
where we also switched to dimensionless variables $x\to (2\pi
T_{c})^2 x$ and $k\to 2\pi T_{c} k$.

By assuming smallness of the critical temperature in comparison 
with the scale put by the magnetic field, $T_c\ell\ll 1$, we see 
that the double logarithmic in field contribution in 
Eq.(\ref{eq:t_c}) comes from the region $0< x \alt 1/2(\pi 
T_{c}\ell)^2 $, $1/\pi \alt k<\infty$.  Simple estimate gives:  
\begin{eqnarray}
1&\simeq&\frac{\alpha}{\pi} \int\limits_0^{1/2(\pi T_{c}\ell)^2} 
dx \int\limits_{1/\pi}^{\infty} \frac{dk}{[1/4 +x]^2 +k^2}\left[
\frac{1/4 +x}{k}+\frac{1/4-x}{\sqrt{k^2+x}}\right]
\nonumber\\
&\simeq& \frac{\alpha}{\pi} \int\limits_0^{1/2(\pi T_c\ell)^2}dx
\left[\frac{1}{2(1/4 +x)} \log\left(1+(1/4 +x)^2\pi^2\right)
     +\frac{1/4-x}{(1/4+x)|1/4-x|}\right.\nonumber\\
& &\left.\times\log\frac{(1/4+x+|1/4-x|)
\sqrt{1/\pi^2+(1/4+x)^2}}{(1/4+x)
\sqrt{1/\pi^2+x}+|1/4-x|/\pi}\right]\simeq 
\frac{\alpha}{4\pi}\log^2\left[\frac{1}{2(\pi T_c\ell)^2}\right].
\label{integ}
\end{eqnarray}
Thus, for the critical temperature, we obtain the estimate:
\begin{eqnarray}
T_{c}\approx \sqrt{|eB|}\exp\left[-\sqrt{\frac{\pi}{\alpha}}\right]
\approx m_{dyn}(T=0),
\label{t_c-m_dyn}
\end{eqnarray}
where $m_{dyn}$ is given by (\ref{massdyn}). The relationship 
$T_c\approx m_{dyn}$ between the critical temperature and the zero 
temperature fermion mass was obtained also in NJL model in (2+1)- 
and (3+1)-dimensions \cite{GMS3,Ebert}.

In passing, let us just briefly note that, the photon thermal mass, 
which is of the order of $\sqrt{\alpha} T$ \cite{Weld}, cannot 
change our result for the critical temperature. As is easy to 
check, the only effect of taking it into account will be the shift
in $x$ for a constant of the order of $\alpha$ in the integrand of 
(\ref{integ}). However, such a shift is absolutely irrelevant for 
our estimate (\ref{t_c-m_dyn}).

In conclusion, we notice that the main result of this paper is the 
analytic estimate for the temperature [given by 
Eqs.(\ref{t_c-m_dyn}) and (\ref{massdyn})] of the chiral symmetry 
restoration in the weakly interacting QED in a background magnetic 
field. In words, the critical temperature is proportional to the 
value of the fermion dynamical mass at zero temperature. Although 
such a result looks natural from the physical point of view, we 
felt necessity to clarify this point by means of a rigorous 
analytical calculation. The reason was the following:  there 
appeared a rough estimate for the critical temperature in 
\cite{Ng2} (such that $T_c\gg m_{dyn}(T=0)$).  This upper estimate 
was enough indeed for the purpose of that paper, i.e.  for proving 
that the catalysis of chiral symmetry breaking is not important in 
dynamics of the electroweak phase transition.  However, making use 
of that rough estimate in other problems may not be appropriate at 
all.

Finally, we would like to emphasize that in this paper as well as 
in all previous publications \cite{QED4,NP,Ng1,Ng2} only was the 
weakly coupled QED ($\alpha\ll 1$) considered. It would also be 
interesting to clarify the influence of an external magnetic field 
on the chiral symmetry breaking in strongly coupled QED ($\alpha > 
\alpha_c$) \cite{RNC} where the chiral symmetry is broken even in 
the absence of a magnetic field. 

\acknowledgments{It is pleasure to thank V.A.~Miransky for 
discussions and helpful comments. We benefited much from 
communications on this subject with Y.J.~Ng whom we would like to 
acknowledge.}


\begin{thebibliography}{99}

\bibitem{GMS1} V.P.~Gusynin, V.A.~Miransky, and I.A.~Shovkovy.
\prl {\bf 73}, 3499 (1994).

\bibitem{Klim} K.G.~Klimenko, Z. Phys. C {\bf 54}, 323 (1992).

\bibitem{GMS2} V.P.~Gusynin, V.A.~Miransky and I.A.~Shovkovy, 
\pl B {\bf 349}, 477 (1995).

\bibitem{GMS3} V.P.~Gusynin, V.A.~Miransky and I.A.~Shovkovy, 
\prd {\bf 52}, 4718 (1995).

\bibitem{QED3} A.V.~Shpagin, {\it Dynamical mass generation in 
(2+1)-dimensional electrodynamics in external magnetic field}, 
hep-ph/9611412 (submitted to Physics of Atomic Nuclei).

\bibitem{QED4} V.P.~Gusynin, V.A.~Miransky and I.A.~Shovkovy, 
\prd {\bf 52}, 4747 (1995).

\bibitem{NP} V.P.~Gusynin, V.A.~Miransky and I.A.~Shovkovy, 
Nucl. Phys. B {\bf 462}, 249 (1996).

\bibitem{Ng1} C.N.~Leung, Y.J.~Ng, and A.W.~Ackley, 
\prd {\bf 54}, 4181 (1996).

\bibitem{Hong} D.K.~Hong, Y.~Kim and S.-J.~Sin, 
\prd {\bf 54}, 7879 (1996).

\bibitem{Igor} I.A.~Shovkovy and V.M.~Turkowski,
\pl B {\bf 367}, 213 (1996).

\bibitem{Ebert} D.~Ebert and V.Ch.~Zhukovsky, {\it Chiral phase 
transitions in strong chromomagnetic fields at finite 
temperature and dymensional reduction}, hep-ph/9701323.

\bibitem{Smilga} I.A.~Shushpanov and A.V.~Smilga, {\it Quark 
condensate in a magnetic field}, preprint ITEP-TH-6/97, 
hep-ph/9703201.

\bibitem{Elias} V.Elias, D.G.C.McKeon, V.A.Miransky, and 
I.A.Shovkovy, \prd {\bf 54}, 7884 (1996).

\bibitem{Ng2}  D.-S.~Lee, C. N.~Leung, and Y. J.~Ng, {\it Chiral 
Symmetry breaking in a uniform external magnetic field}, 
hep-th/9701172 (to appear in Phys.Rev D, May issue, 1997).

\bibitem{RNC} P.I.~Fomin, V.P.~Gusynin, V.A.~Miransky, and 
Yu.A.~Sitenko, Riv. Nuovo Cim. {\bf 6}, N5 (1983).

\bibitem{Jancov} B.~Jancovici, Phys. Rev. {\bf 187}, 2275 (1969).

\bibitem{Loskutov} Yu.M.~Loskutov and V.V.~Skobelev, Theor. Mat. 
Fiz. {\bf 48}, 44 (1981).

\bibitem{Abr} A.A.~Abrikosov, L.P.~Gorkov and I.E.~Dzyaloshinski, 
{\it Methods of Quantum Field Theory in Statistical Physics} 
(Nauka, Moscow, 1962).

\bibitem{Weld} H.A.~Weldon, \prd {\bf 26}, 1394 (1982).

\end{thebibliography}
\end{document}